\documentclass[twocolumn,superscriptaddress,showpacs,preprintnumbers,amsmath,amssymb]{revtex4}
\usepackage{amsmath}
\usepackage{amssymb}
\usepackage{dcolumn}% Align table columns on decimal point
\usepackage{graphicx}
\usepackage{bm}% bold math

\begin{document}
%\begin{CJK}{UTF8}{gbsn} % Use default fonts from CJK (see below)
\title{Can the packing efficiency of binary hard spheres explain the
glass-forming ability of bulk metallic glasses?} 
\author{Kai Zhang} 
%(张凯)}
%\email{}
\affiliation{Department of Mechanical Engineering and Materials Science, Yale University, New Haven, Connecticut, 06520, USA}
\affiliation{Center for Research on Interface Structures and Phenomena, Yale University, New Haven, Connecticut, 06520, USA}
\author{W. Wendell Smith}
%\email{}
\affiliation{Department of Physics, Yale University, New Haven, Connecticut, 06520, USA}
\author{Minglei Wang} % (王明磊)}
%\email{}
\affiliation{Department of Mechanical Engineering and Materials Science, Yale University, New Haven, Connecticut, 06520, USA}
\affiliation{Center for Research on Interface Structures and Phenomena, Yale University, New Haven, Connecticut, 06520, USA}

\author{Yanhui Liu} % (柳延辉)}
\affiliation{Department of Mechanical Engineering and Materials Science, Yale University, New Haven, Connecticut, 06520, USA}
\affiliation{Center for Research on Interface Structures and Phenomena, Yale University, New Haven, Connecticut, 06520, USA}
\author{Jan Schroers}
\affiliation{Department of Mechanical Engineering and Materials Science, Yale University, New Haven, Connecticut, 06520, USA}
\affiliation{Center for Research on Interface Structures and Phenomena, Yale University, New Haven, Connecticut, 06520, USA}
\author{Mark D. Shattuck}
\affiliation{Department of Physics and Benjamin Levich Institute, The City College of the City University of New York, New York, 10031, USA}
\affiliation{Department of Mechanical Engineering and Materials Science, Yale University, New Haven, Connecticut, 06520, USA}
\author{Corey S. O'Hern}
\affiliation{Department of Mechanical Engineering and Materials Science, Yale University, New Haven, Connecticut, 06520, USA}
\affiliation{Center for Research on Interface Structures and Phenomena, Yale University, New Haven, Connecticut, 06520, USA}
\affiliation{Department of Physics, Yale University, New Haven, Connecticut, 06520, USA}
\affiliation{Department of Applied Physics, Yale University, New Haven, Connecticut, 06520, USA}

\date{\today}

\begin{abstract}

We perform molecular dynamics simulations to compress binary hard
spheres into jammed packings as a function of the compression rate
$R$, size ratio $\alpha$, and number fraction $x_S$ of small particles
to determine the connection between the glass-forming ability (GFA)
and packing efficiency in bulk metallic glasses (BMGs).  We define the
GFA by measuring the critical compression rate $R_c$, below which
jammed hard-sphere packings begin to form ``random crystal''
structures with defects.  We find that for systems with $\alpha
\gtrsim 0.8$ that do not de-mix, $R_c$ decreases strongly with $\Delta
\phi_J$, as $R_c \sim \exp(-1/\Delta \phi_J^2)$, where $\Delta \phi_J$
is the difference between the average packing fraction of the
amorphous packings and random crystal structures at $R_c$.  Systems
with $\alpha \lesssim 0.8$ partially de-mix, which promotes
crystallization, but we still find a strong correlation between $R_c$
and $\Delta \phi_J$. We show that known metal-metal BMGs occur in the
regions of the $\alpha$ and $x_S$ parameter space with the lowest
values of $R_c$ for binary hard spheres. Our results emphasize that
maximizing GFA in binary systems involves two competing effects:
minimizing $\alpha$ to increase packing efficiency, while maximizing
$\alpha$ to prevent de-mixing.
\end{abstract}

\pacs{64.70.pe,%metallic glasses
64.70.Q-,%theory and modeling of the glass transition
61.43.Fs,%glasses
61.66.Dk,%alloys
} \maketitle

%\end{CJK}

Hard-sphere models provide quantitatively accurate descriptions of
physical properties in systems where steric, repulsive
interactions are dominant, such as the diverging viscosity near the glass
transition in colloids~\cite{tanaka}, transport properties in simple
liquids~\cite{hansen}, and mechanical properties of granular
materials~\cite{makse}.  For more complex materials with competing
repulsive and attractive interactions, such as bulk metallic glasses,
it is often helpful to develop a perturbative description where only
hard-sphere interactions~\cite{wca} are included to determine to what extent
these alone can explain key physical properties~\cite{miracle_2004}.

Bulk metallic glasses (BMGs) are prepared by thermally quenching
liquid alloys at sufficiently fast rates such that they bypass
crystallization, and instead form amorphous
solids~\cite{inoue:2000,wang_2004}.  Over the past $30$ years, BMGs
have been developed with optimized mechanical properties, such as
enhanced strength and fracture toughness above that for
steel~\cite{schuh_2007}, but with processing and molding capabilities
similar to plastics~\cite{schroers}. However, their applications in
industry are still often constrained by the high cost of the
constituent elements and the maximum casting thickness of the
material.  The glass-forming ability (GFA) of a BMG is defined by the
critical cooling rate below which the system begins to crystallize,
which in turn, determines its critical casting
thickness~\cite{inoue:2000}.  An important open question is how
to {\it de novo} design BMGs with desirable material properties and
maximum glass formability by continuously varying the stoichiometry of
the constituent elements~\cite{progress}.

There are well-known empirical rules for improving the glass-forming
ability of BMGs, for example, increasing the number of components, and
ensuring that the atomic size difference, for at least some of the
constituents, is above $12\%$ and that the heats of mixing among the
main constituent elements are negative~\cite{inoue}. A number of more
recent studies have identified quantities that are correlated with
GFA, such as the viscosity~\cite{wang_2012}, glass transition and
crystallization temperatures~\cite{lu_2002,du_2007},
atomic~\cite{miracle_2010,cheng:2009} and electronic~\cite{yu_2010}
structure.  Despite these guiding principles, we still lack a
predictive understanding of BMG formation.  For example, we do not
even know the relative entropic and enthalpic contributions to the
glass-forming ability of metal alloys, which would be a first step in
computationally designing new BMGs with arbitrary compositions.

\begin{figure*}%[!b]
\includegraphics[width=3.5in]{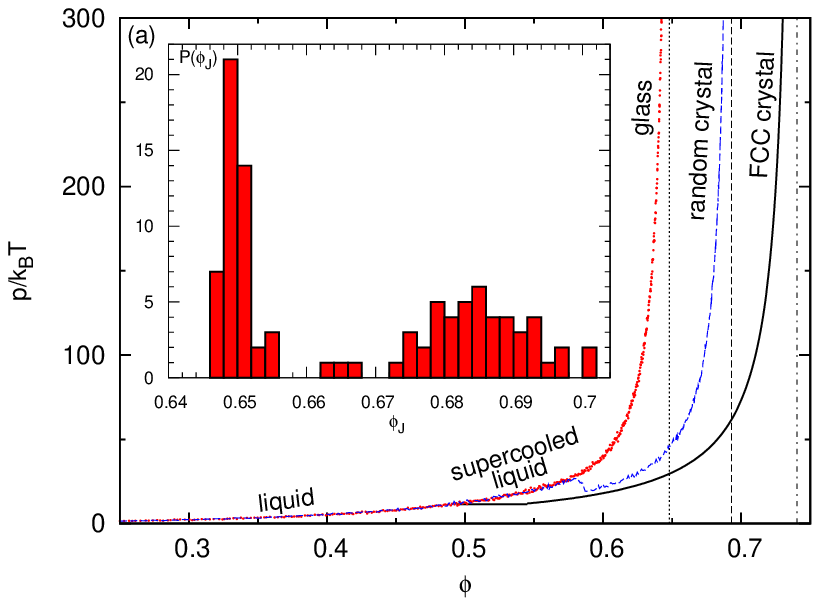}
\includegraphics[width=3.5in]{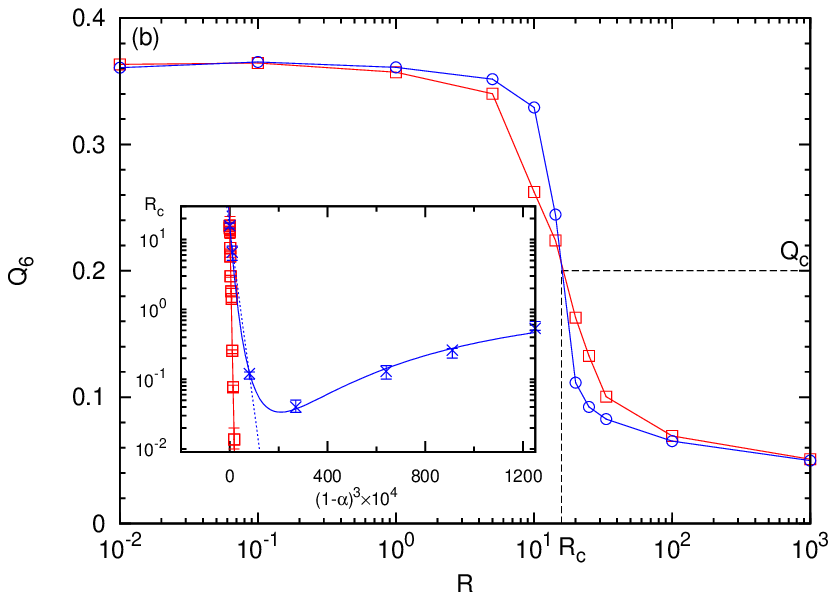}
\caption{(Color online) (a) Pressure $p/k_B T$ versus packing fraction
$\phi$ for monodisperse hard spheres at compression rate $R \sim R_c
\approx 10$.  The red dotted and blue dashed branches terminate at
$\phi_{J}^a \approx 0.648$ (vertical dotted line) and $\phi_{J}^x
\approx 0.693$ (vertical dashed line), which correspond to typical
disordered and ``random crystal'' configurations. The equilibrium
$p(\phi)$ (black solid line) terminates at the close packed face centered 
cubic (FCC) crystal with $\phi_{c} = \pi/\sqrt{18}$~\cite{hoste:1984} (vertical
dot-dashed line).  In the inset, we show the probability distribution
$P(\phi_J)$ of jammed packing fractions from $96$ random initial
conditions.  (b) The mean (squares) and median (circles) global
bond-orientational order parameter $Q_6$ versus $R$ for $\alpha=1$. We
define the critical compression rate $R_c$ (and $Q_c$) by the
intersection of the mean and median $Q_6$. In the inset, we show that
for $x_S = 0.5$, $R_c$ is monotonic over the given dynamic range and
scales as $R_c\sim \exp[-C (1-\alpha)^3]$, with $C\approx 4000$ (solid
line), for $0.88 < \alpha < 1$.  $R_c$ obeys similar scaling
for $x_S = 0.2$ (with $C\approx 600$; dotted line), but $R_c$ begins
to increase for $\alpha < 0.8$.}
\label{fig:hspphi}
\end{figure*}

We focus on a simple model glass-forming system,
bidisperse hard spheres, to quantify the entropic contribution to the
glass-forming ability as a function of the atomic size ratio $\alpha$,
number fraction $x_S$ of small atoms, and compression rate $R$ (which
is analogous to the cooling rate in systems with soft interaction
potentials). When hard-sphere systems are compressed sufficiently
rapidly, they do not undergo an equilibrium freezing transition, and
remain structurally disordered on the metastable branch of the
equation of state.  Upon further compression, hard-sphere systems jam
into one of many packings with vanishing free volume at packing
fraction $\phi_J$, which depends on the compression rate as well as
the initial condition as shown in Fig.~\ref{fig:hspphi} (a).  In the
infinite compression rate limit, $\phi_J(\infty)$ approaches random
close packing (with $\phi_{\rm rcp} \approx 0.64$ for monodisperse
spheres~\cite{rintoul:1996}).  In the $R \rightarrow 0$ limit,
hard-sphere packings form perfect crystalline structures at $\phi_c >
\phi_J$ (with face-centered cubic symmetry and $\phi_c = \pi/\sqrt{18}$ for
monodisperse spheres).  At finite, but slow compression rates,
``random crystals'' form with many crystal defects and
amorphous domains with $\phi_{\rm rcp} < \phi_{J}^x < \phi_c$.

\begin{figure}%[!b]
\includegraphics[width=3.6in]{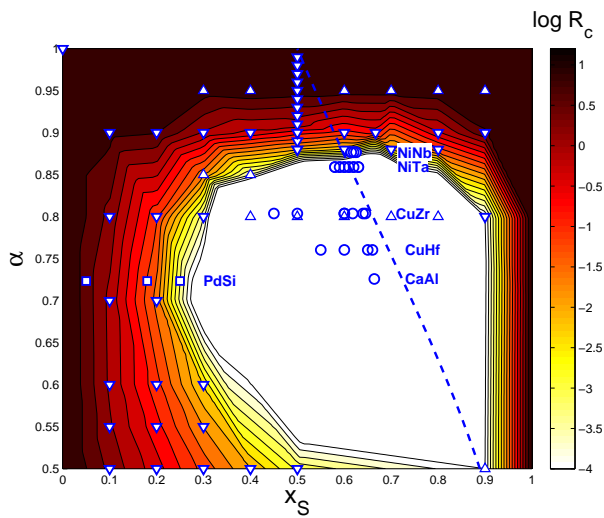}
\caption{(Color online) Contour plot of $R_c$ versus $\alpha$ and
$x_S$. The shading from dark to light indicates decreasing $R_c$ on a
logarithmic scale. The downward triangles are from MD simulations, the
upward triangles are obtained by fitting $R_c$ to Eq.~\ref{rc}, and
the circles and squares correspond to known metal-metal ({\it e.g.}
NiNb, NiTa, CuZr, CuHf, and CaAl~\cite{xu:2004,xia:2006,wang:2010})
and metal-metalloid ({\it e.g.}  PdSi~\cite{lu:2002}) binary BMGs,
respectively.  The dashed line satisfies $x_{S}^* =
(1+\alpha^3)^{-1}$, at which the large and small particles occupy the
same volume.}
\label{fig:RcContour}
\end{figure}

We seek to determine the variables that control the critical
compression rate $R_c$, below which crystalline domains begin to form
in bidisperse hard-sphere systems as a function of $\alpha$ and
$x_S$. For example, is the packing fraction of crystalline
configurations with a particular type of order important, and if so,
which one at each $\alpha$ and $x_S$?  Or is the packing fraction of
typical amorphous configurations more important for determining the
glass formability?  Our computational studies show that over a wide
range of size ratios and compositions where partial de-mixing does not
occur, $R_c \sim \exp(-1/\Delta \phi_J^2)$ is controlled by the
packing fraction difference $\Delta \phi_J$ between the average
packing fraction of the amorphous configurations and that of the
competing ``random crystal'' configurations.  For systems with $\alpha
< \alpha_c$, partial de-mixing intervenes and $R_c$ has a more complex
dependence on $\Delta \phi_J$.  Further, we show that most known
metal-metal binary bulk metallic glasses occur in the region of the
$\alpha$ and $x_S$ parameter space with the smallest $R_c$ for
bidisperse hard-sphere mixtures (Fig.~\ref{fig:RcContour}), which
suggests that the hard-sphere model is sufficient for explaining
important general features of the GFA of metal-metal BMGs.
   
We study binary hard-sphere mixtures of $N=N_L+N_S=500$ particles with
the same mass $m$ and diameter ratio $\alpha = \sigma_S/\sigma_L < 1$
of small to large particles using event-driven molecular dynamics (MD)
simulations within a cubic box of volume $V$ under periodic boundary
conditions.  We first prepare equilibrium liquids at a given $\alpha$
and small particle fraction $x_S=N_S/N$ at initial packing fraction
$\phi = \frac{\pi}{6}\frac{N \sigma_L^3}{V}\left(1 +
(\alpha^3-1)x_S\right) = 0.25=\phi_0$.  To compress the system, we
increase the particle sizes by a factor $\gamma=\underset{i<j}{\min}
\{r_{ij}/\sigma_{ij}\}$, while preserving $\alpha$, until the first
pair of spheres comes into contact~\cite{jalali:2004,jalali:2005},
where $r_{ij}$ is the separation between particles $i$ and $j$ and
$\sigma_{ij}= (\sigma_{i} + \sigma_{j})/2$.  Between each compression,
the system is equilibrated at constant volume for a time interval
$\tau$, during which we measure the collision
frequency and pressure.  In the supplementary material, we show that
this protocol gives rise to an exponential approach to the final
jammed packing fraction $\phi_{J}$: $\phi_{J} - \phi(t) = (\phi_{J} -
\phi_0) e^{-{\widetilde R}t}$, where ${\widetilde R} = k R$, $k$ is a
constant, and $R=1/\tau$ (expressed in units of $\sqrt{k_B T/m
\sigma_L^2}$, where $k_B T$ is the thermal energy) is used to vary the
compression rate. We terminate the hard-sphere MD compression protocol
when the pressure exceeds $p/k_BT = 10^3$ at $\phi_J'$. We then
implemented soft-particle techniques~\cite{gao} to compress the
packings at $p/k_BT \sim 10^3$ to jammed packings at $p \rightarrow
\infty$, $\phi_J >\phi_J'$, with $(\phi_J -\phi_J')/\phi_J' \ll 1$.

At each $R$, we compress $96$ systems with different random initial
particle positions to generate the distribution $P(\phi_J)$ of jammed
packing fractions. As shown in the inset of Fig.~\ref{fig:hspphi} (a),
$P(\phi_J)$ is bimodal with a narrow peak corresponding to
amorphous configurations and a broad peak corresponding to random
crystal configurations.  We also calculate the global
bond-orientational order parameter $Q_6$ for each
configuration~\cite{steinhardt:1983}, where nearest neighbor particles
are identified using Voronoi
tessellation~\cite{sheng:2006}. We find that $P(Q_6)$ also
exhibits a bimodal distribution as shown in the supplementary
material.

The relative weight of the two peaks in $P(Q_6)$ shifts toward the
random crystal peak as $R$ decreases, which causes both the mean and median
$Q_6$ to increase.  We define the critical compression rate $R_c$ at
the intersection of the mean and median $Q_6$ (Fig.~\ref{fig:hspphi}
(b)). We then measured $R_c$ versus $\alpha$ and $x_S$ and
found several key results.  First, for relatively large $\alpha \sim
1$, $R_c$ decreases exponentially as 
\begin{equation}
\label{rc}
R_c\sim \exp[-C(1-\alpha)^3],
\end{equation}
where $C$ depends on $x_S$~\cite{zhang:2013}. As $\alpha$ decreases
further, $R_c$ becomes nonmonotonic, as shown in the inset to
Fig.~\ref{fig:hspphi} (b) for $x_S=0.2$. Second,
most known binary bulk metallic glass-forming alloys
possess $\alpha$ and $x_S$ with the smallest values of $R_c$ for
binary hard-spheres. In Fig.~\ref{fig:RcContour}, we show a contour
plot of $R_c$ as a function of $\alpha$ and $x_S$ for binary hard
spheres. To construct the contours, we directly measured $R_c$
(downward triangles) from MD simulations as well as employed
Eq.~\ref{rc} to extrapolate $R_c$ (upward triangles) in systems where
$R_c < 10^{-3}$ is below the simulation threshold. We identify a
region bounded approximately by $0.45 \lesssim \alpha \lesssim 0.85$
and $0.35 \lesssim x_S \lesssim 0.9$ where the binary hard-sphere
model predicts $R_c \lesssim 10^{-4}$.  Note that for $\alpha < 0.7$,
the good glass-forming regime shifts toward increasingly larger
$x_S$. In contrast, the good glass-forming regime near $\alpha = 0.85$
includes the broadest range of $x_S$. Our previous studies of binary
Lennard-Jones~\cite{zhang:2013} and current studies of hard-sphere
mixtures (supplementary materials) show that the composition with the
smallest $R_c$ at each $\alpha$ is $x_S^* = 1/(1+\alpha^3)$ at which
the large and small particles occupy the same volume.  Binary
metal-metal BMGs~\cite{xu:2004,xia:2006,wang:2010} tend to cluster
near $x_S^*$ and populate the low-$R_c$ region of the contour plot. In
contrast, binary metal-metalloid BMGs~\cite{lu:2002} do not cluster
near $x_S^*$, possess only a small fraction ($\lesssim 30 \%$) of
small atoms, and lie outside the low-$R_c$ region for binary hard
spheres~\cite{guan_2012}.  

\begin{figure}%[!b]
\includegraphics[width=3.5in]{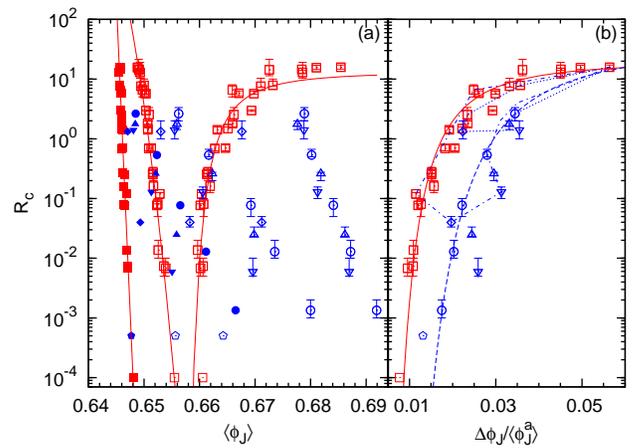}
\caption{(Color online) (a) $R_c$ and the corresponding packing
fractions $\langle \phi^a_J \rangle$ and $\langle \phi^x_J \rangle$
evaluated at $R_c$ (open symbols with $\langle \phi^a_J \rangle <
\langle \phi^x_J \rangle$). $\langle \phi_J^a \rangle$ for jammed
packings in the $R \rightarrow \infty$ limit (filled symbols) are also
shown.  We compare $\langle \phi^a_J \rangle$ and $\langle \phi^x_J
\rangle$ for systems with $\alpha \gtrsim 0.8$ (red squares), which
remain well-mixed even after forming random crystals ({\it i.e.}
polymorphic crystallization), and systems with $\alpha = 0.8$ and
$x_S=0.3$ (pentagons) as well as $\alpha=0.7$ (diamonds), $0.6$
(downward triangles), $0.55$ (upward triangles), and $0.5$ (circles)
over a range of $x_S$, which partially de-mix before (non-polymorphic)
crystallization. (b) $R_c$ versus packing fraction deviation $\Delta
\phi_J = \langle \phi^x_J \rangle - \langle \phi^a_J \rangle$. Systems
that remain well-mixed (squares) collapse onto the master curve given
by Eq.~\ref{log_rc} (solid line) with $-a \approx 4\times 10^{-4}$ and
$I_{\infty} \approx 1.3$. We highlight systems at fixed composition,
$x_S=0.1$ (dotted line) and $0.2$ (dot-dashed line), and varying
$\alpha$. The dashed line shows Eq.~\ref{log_rc} with $-a \approx
10^{-3}$ and $I_{\infty} \approx 1.6$, which fits the $R_c$ data for
$\alpha=0.5$. Error bars give the standard deviation over $96$ initial
conditions.}
\label{fig:Rcphi}
\end{figure}

We now seek to determine the connection between the glass-forming
ability measured by $R_c$ and packing efficiency by focusing on the
mean packing fractions $\phi_J^a$ and $\phi_J^x$ of the subpopulations of
amorphous and random crystal configurations, respectively, at
$R_c$. To calculate $\langle \phi_J^a \rangle$ ($\langle \phi_J^x
\rangle$), we average the packing fractions of the jammed
configurations with $Q_6<Q_c$ ($Q_6 > Q_c$).  In Fig.~\ref{fig:Rcphi}
(a), we plot $R_c$ and the corresponding $\langle \phi_J^a \rangle$
and $\langle \phi_J^x \rangle$ for each $\alpha$ and $x_S$ pair
studied. We find that (for systems that remain well mixed) decreases
in $R_c$ are accompanied by increases in the packing efficiency of the
amorphous configurations and decreases in the random crystal packing
efficiency.  We can identify a relation between $R_c$ and the packing
fraction deviation $\Delta \phi_J \equiv \langle \phi_J^x \rangle -
\langle \phi_J^a \rangle$ by comparing $R_c$ and the nucleation rate,
$I$,
\begin{equation}
R_c  \sim  I = I_{\infty} e^{-\Delta G^*/k_BT},
\end{equation}
where $\Delta G^* \sim \gamma^3/\Delta \mu^2$ is the nucleation free
energy barrier, $\gamma$ is the surface tension of random crystal
clusters, and $\Delta \mu$ is the volume contribution to the change in
free energy from adding a particle to  cluster, and $I_{\infty}$ is the
kinetic prefactor. For hard spheres, $\Delta \mu = -k_B T \Delta S$, 
$\log R_c \sim 1/\Delta S^2$, and thus
\begin{equation}
\label{log_rc}
\log R_c = a  (\Delta \phi_J/\langle \phi_J^a \rangle)^{-2} + \log I_{\infty}, 
\end{equation}
where $a<0$, for $\Delta \phi_J/\langle \phi^a_J \rangle \ll 1$. The
thermodynamic drive for random crystal formation scales to zero with $\Delta
\phi_J$, which enhances the glass formability. We show in
Fig.~\ref{fig:Rcphi} (b) that Eq.~\ref{log_rc} collapses the data for
$R_c$ for $\alpha \gtrsim 0.8$.  However, for systems with $\alpha
\lesssim 0.8$, the behavior of $R_c$ is more complicated.

\begin{figure}%[!b]
\includegraphics[width=3.5in]{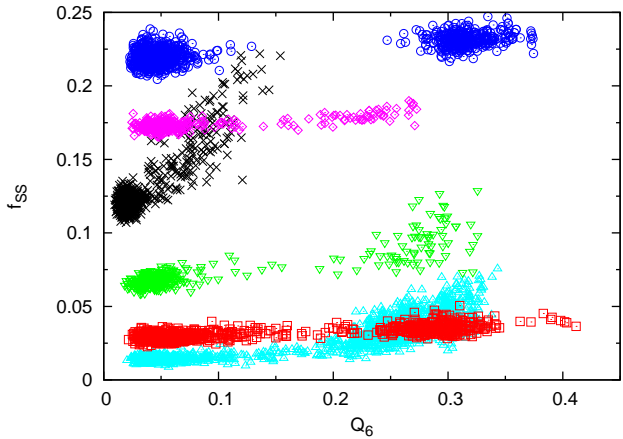}
\caption{(Color online) Fraction of small-small nearest neighbors
$f_{SS}$ versus $Q_6$ for six $(\alpha,x_S)$ values: $(0.5,0.5)$
(crosses), $(0.5,0.2)$ (upward triangles), $(0.8,0.3)$ (downward
triangles), $(0.8,0.9)$ (diamonds), $(0.8,0.2)$ (squares), and
$(0.88,0.5)$ (circles). $f_{SS}$ for $(0.8,0.9)$ has been shifted
downward by $0.6$ to enable comparison with the other systems. 
The systems with $(0.5,0.5)$, $(0.5,0.2)$, and $(0.8,0.3)$ partially de-mix 
as indicated by the increase in $f_{SS}$ with increasing $Q_6$.}
\label{fig:fss}
\end{figure}

In Fig.~\ref{fig:fss}, we show the fraction of small-small nearest
neighbors~\cite{schreck:2011} as a function of $Q_6$ for several
$\alpha$ and $x_S$ pairs, where nearest neighbor particles share
Voronoi polyhedra faces. We find that systems with $\alpha \lesssim
0.8$ exhibit partial de-mixing prior to the formation of random
crystals ({\it i.e.} non-polymorphic crystallization) as evidenced by
the increase in $f_{SS}$ with increasing $Q_6$. For systems with small
size ratios, {\it e.g.} $\alpha=0.5$, the large particles form the
rigid backbone of the random crystal, while the small particles, which
can fit in the interstices of the large-particle backbone, remain
disordered. We find that de-mixing encourages the formation of random
crystals, which results in lower $R_c$ at the same $\Delta \phi_J$
compared to systems that remain well-mixed.  Even though there is more
scatter for the systems that partially de-mix, $R_c$ decreases
strongly with decreasing $\Delta \phi_J$ as $x_S$ is varied at fixed
$\alpha$. At fixed composition ({\it e.g.}  $x_S=0.1$ or $0.2$), $R_c$
versus $\Delta \phi_J$ deviates from the $\alpha \gtrsim 0.8$ master
curve as $\alpha$ decreases below $0.8$, but it eventually reconnects
with the monodisperse systems for sufficiently small $\alpha$.

Structural differences between the well-mixed and de-mixed systems can
also be found in the disordered configurations in the $R \rightarrow
\infty$ limit. For example, $\langle \phi_J^a \rangle$ obtained from
jammed packings in the $R \rightarrow \infty$ limit is strongly
correlated with $R_c$ for the well-mixed systems, however, the data is
highly scattered for the de-mixed systems. By analyzing the radial
distribution function $g(r)$, we find that the structural symmetry
between the small and large particles does not occur for $\alpha
\lesssim 0.8$. Instead, the large particles form the rigid backbone of
the jammed packing, while the peaks in $g(r)$ corresponding to
separations between small particles broaden and become
liquid-like, as shown in the supplementary material.

In conclusion, we have shown that the binary hard-sphere model
explains several general features of the GFA for metal-metal BMGs. In
particular, we find that for systems with $\alpha \gtrsim 0.8$ that do
not de-mix, $R_c \sim \exp(-1/\Delta \phi_J^2)$ is set by the average
packing fraction deviation $\Delta \phi_J$ between the amorphous and
random crystal configurations, and $R_c \rightarrow 0$ as $\Delta
\phi_J$ tends to zero. For systems with $\alpha \lesssim 0.8$ that
partially de-mix, each $R_c(\alpha)$ obeys a similar curve. In addition,
most known metal-metal BMGs occur in the low-$R_c$ region of $\alpha$
and $x_S$ parameter space for binary hard-spheres, but metal-metalloid
BMGs do not. Our studies show that maximizing the glass-forming
ability in binary systems involves competing effects: minimizing
$\alpha$ to increase packing efficiency and maximizing $\alpha$ to
reduce the tendency for de-mixing. This suggests that the GFA 
can be increased in ternary systems by preventing de-mixing. 

\begin{acknowledgments}
We acknowledge primary financial support from the National Science
Foundation (NSF) MRSEC DMR-1119826 (K.Z.) and partial support from NSF
Grant Nos.  DMR-1006537 (C.S.O.) and CBET-0968013 (M.D.S.). This work
also benefited from the facilities and staff of the Yale University
Faculty of Arts and Sciences High Performance Computing Center and the
NSF (Grant No. CNS-0821132) that in part funded acquisition of the
computational facilities.
\end{acknowledgments}

\end{document}